\documentstyle[epsfig,prl,aps,preprint]{revtex}

\begin{document}
\draft \title {Normal and Anomalous Diffusion in a Deterministic 
Area-preserving Map}
\author{{\large P. Leb{\oe}uf}}
\address{Division de Physique Th\'eorique\footnote{Unit\'e de recherche 
des Universit\'es de Paris XI et Paris VI associ\'ee au CNRS}, Institut de 
Physique Nucl\'eaire, 91406 Orsay Cedex, France}

\maketitle
\begin{abstract}
Chaotic deterministic dynamics of a particle can give rise to diffusive
Brownian motion. In this paper, we compute analytically the diffusion
coefficient for a particular two-dimensional stochastic layer induced by the
kicked Harper map. The variations of the transport coefficient as a parameter
is varied are analyzed in terms of the underlying classical trajectories with
particular emphasis in the appearance and bifurcations of periodic orbits. When
accelerator modes are present, anomalous diffusion of the L\'evy type can occur.
The exponent characterizing the anomalous diffusion is computed numerically and 
analyzed as a function of the parameter.
\end{abstract}

\vspace{3cm}

\noindent IPNO/TH 96-50 ~~~~~~~~~~~~~~~~~~~~~~~~~~~~~~~~~~~~~~~~~~~~~~~~~~~~~
~~~~~~~~~~~~~~~~~~~~~~~~~~~~~ Novembre 1996\\
\pagebreak

\narrowtext

\section{Introduction}

Transport properties of deterministic chaotic Hamiltonian dynamics can exhibit 
a rich variety of phenomena. In many situations where chaos is strong, 
transport through phase space tends to be uniform and can imitate random 
processes of the Brownian type. The motion on average is a normal diffusion.
However, in more generic situations where regular as well as chaotic behaviour 
coexist the phase space has a fractal structure, and this structure can 
induce important variations of the transport process.

In this paper our purpose is to investigate transport properties of a 
deterministic two-dimensional area preserving map whose dynamics can be either 
mixed or strongly chaotic as a parameter is varied. It is defined by the
time-dependent Hamiltonian 
\begin{equation} \label{1}
\begin{array}{ccl} 
H(q,p,t)&=&-V_2\cos(2\pi p) - V_1\cos(2\pi q)\,K(t) \\
K(t)&=&\tau\sum_{n=-\infty}^\infty\delta(t-n\tau)
\end{array}
\end{equation}
called the kicked Harper model. It arises in several branches of physics, as
for example in the description of charged particles interacting with an
electrostatic wave packet in a transverse uniform magnetic field
\cite{zzsuc,zas,dana}, in the study of the two-dimensional motion of electrons
in a periodic potential plus uniform perpendicular magnetic field \cite{lkfa}
and in connection to the steady-state dynamics of two and three-dimensional
ideal incompressible fluids \cite{zsc,pas}. A characteristic feature of
Harper-like Hamiltonians is the particular functional form of the kinetic
energy. The equations of motion induced by Eq.(\ref{1}) can be integrated over
one period $\tau$ of the kicking potential, leading to the map 
\begin{equation} \label{2}
\begin{array}{ccc} 
p_{n+1}&=&p_n-\gamma_1\sin(2\pi q_n) \\
q_{n+1}&=&q_n+\gamma_2\sin(2\pi p_{n+1}) 
\end{array}
\end{equation}
where $\gamma_i=2\pi V_i\tau$. We can consider the motion either in the entire 
phase-space plane or in the fundamental domain $(q,p)\in [0,1[^2$. The 
existence of a fundamental domain comes from the fact that the map (\ref{2}) 
is invariant under translations $(q,p)\rightarrow (q+l,p+m)$ -- where $(l,m)$ 
are two arbitrary positive or negative integers -- since such a shift of the 
initial conditions produces a global shift of the trajectory.

In the limit $\tau\to 0$ the model reduces to the continuous time evolution of
the Harper Hamiltonian $H_0 (q,p)=-V_2 \cos(2\pi p)- V_1 \cos(2\pi q)$ which,
from the dynamical point of view, is an integrable system. For simplicity, we
restrict the following discussion to the special case $\gamma_1 = \gamma_2 =
\gamma$ ($V_1 = V_2$). Then the phase space plots of $H_0$ are characterized by
the existence of a separatrix defined by $H_0 (q,p) = 0$ connecting both sides
of the fundamental domain.  In the full phase plane, the separatrix forms a
regular lattice having a square symmetry (cf Fig.1(a)), while the contour lines
$H_0 (q,p) = {\rm const} \neq 0$ are closed curves. For $\gamma> 0$ the
separatrix is destroyed and becomes a mesh of finite thickness inside which the
dynamics is chaotic. This chaotic mesh, which has been called the stochastic
web \cite{zzsuc}, coexists with regions of phase space where regular motion takes
place. As $\gamma$ increases the stochastic web grows, and for sufficiently
large values of $\gamma$ the web disappears and chaotic dynamics covers most of 
phase space. This transition is shown in Fig.1 for the motion restricted to the
fundamental domain. In order to better visualize the geometry of the stochastic
web, in Fig.2 we have plotted again Fig.1(b) but now in the full phase plane. 

The motion of the particle is not chaotic for any initial condition inside the 
stochastic web, since there are regular islands on it (clearly visible in 
Fig.2). However, if an initial condition corresponding to a chaotic motion on 
the web is chosen then, because the web extends over the whole phase-space 
plane, the particle can "diffuse" arbitrarily far from its initial position. 
In the $p$ direction, the diffusion will be characterized by the coefficient 
\begin{equation} \label{3}
D=\lim_{n\rightarrow\infty} \frac{\langle (p_n-p_0)^2 \rangle_R}{n}
\end{equation}
where the average is taken over a set $R$ of phase-space initial conditions.
This coefficient can in general be expressed as an infinite series involving
the force-force correlation functions (see \cite{cm} and Eq.(\ref{7}) below).
The existence of $D$ (i.e., the convergence of the series) and therefore the
existence of a true deterministic diffusive process is in general difficult to
prove rigorously. This, however, has been done for a different map, the
sawtooth map \cite{cm}. Moreover, several aspects of the diffusion process in
deterministic maps have been studied in the past by truncation of the series
and/or adding some noise. These studies were mainly concentrated on the
standard map \cite{chiri,rw,abar} or on different types of stochastic webs
\cite{zzsuc,lw}. In particular, computations of $D$ for the kicked Harper map
can be found in \cite{acsz,da}. Our aim here is to compute the diffusion
coefficient as a function of $\gamma$ up to second order in the correlation
functions and to analyze its dependence on $\gamma$ in terms of the underlying
classical dynamics (section III). We obtain, for all $\gamma$, a finite value
of $D$ which is in very good agreement with the numerical results except near
some parameter ranges where accelerator modes appear. These modes are stable
periodic orbits which coherently propagate across the web. The shortest
periodic orbits corresponding to accelerator modes as well as their stability
are computed in Section II. The existence of these orbits produces a divergence
of the diffusion coefficient if the average in Eq.(\ref{3}) is taken over all
possible initial conditions. This is because in the presence of accelerator
modes the average can be splitted into two parts: the first is over the stable
phase-space regions  ${\cal S}$ where accelerator modes exist, while the second
is over the chaotic component ${\cal C}$ 
\begin{equation} \label{3p}
\langle (p_n-p_0)^2 \rangle_{R_f} = \langle (p_n-p_0)^2 \rangle_{\cal S} +
\langle (p_n-p_0)^2 \rangle_{\cal C} \ .
\end{equation}
where the index $R_f$ means that the average has been taken over the whole
fundamental domain. Because accelerator modes propagate in a ballistic manner,
$\langle (p_n-p_0)^2 \rangle_{\cal S} \sim n^2$ and the diffusion coefficient
associated to this region diverges like $n$.

The presence of accelerator modes can have more unexpected consequences on the
transport properties of the system than the one just mentioned. Their influence
extends into the chaotic sea and may produce non-local effects. In fact, the
partial barriers (cantori) induced by their presence in the chaotic regions of
the web may modify the nature of the transport in the stochastic web as well.
For most of the time the particle diffuses, but sometimes it is trapped in the
neighbourhood of an accelerator modes and drag along in a ballistic manner.
These rare events can dominate the dynamics and produce anomalous diffusion of
the L\'evy type \cite{levy} 
\begin{equation} \label{4}
\langle (p_n-p_0)^2 \rangle_{\cal C} \sim n^\beta, 
\;\;\;\;\;\;\;\; 1<\beta< 2 \ .
\end{equation}
As a consequence the diffusion coefficient associated to the chaotic component
also has a power-law divergence with exponent $\beta-1$. 

Anomalous diffusion was already observed numerically in different Hamiltonian
systems, like the standard map \cite{karney,cs,ihhm,zk}, in stochastic webs
\cite{cprsz,asz} and in two-dimensional smooth potentials \cite{cz}. There are
however few systematic quantitative results. At present there is no theory
making the connexion between the existence of phase space barriers in the
chaotic layers on the one hand and anomalous diffusion of the L\'evy type on
the other hand. Only in one case it was possible to establish \cite{gt} a
direct connexion between some parameter of a map and the temporal exponent
$\beta$ determining the anomalous diffusion. But unfortunately this was done in
the simple case of a one-dimensional iterative map. In Sect.III we compute, as
a function of $\gamma$, the exponent $\beta$. Our numerical results, presented
in Fig.5 and Fig.6, demonstrate that the kicked Harper map provides an
additional example of anomalous diffusion of the L\'evy type in Hamiltonian
systems. Different aspects of this exponent in connexion to the existence of 
normal modes are discussed. 

\section{Accelerator modes}

The period-$1$ fixed points of the map (\ref{2}) are determined by the 
condition 
$p_{n+1}=p_n$, $q_{n+1}=q_n$. We are interested however in the fixed points 
of the map on the fundamental domain, and thus we identify phase-space points 
which differ from an integer value: $p_{n+1}=p_n+w_p$, $q_{n+1}=q_n+w_q$
leading to
\begin{equation} \label{5}
\begin{array}{ccc} 
\sin(2\pi p_0)&=&-w_q/\gamma_1 \\
\sin(2\pi q_0)&=&-w_p/\gamma_2 
\end{array}
\end{equation}
where $(w_q,w_p)\in Z^2$ and we have used the notation $q_n=q_0$, $p_n=p_0$.
These two integers are the winding numbers of the period-$1$ fixed points since
they define the number of turns the orbit does around the fundamental domain in
one iteration of the map. In the phase plane, periodic orbits with winding
numbers $(w_q,w_p)$ propagate ballistically, with $(p_n-p_0)=w_p n$ and
$(q_n-q_0)=w_q n$. But not all of them correspond to accelerator modes either
because they are not stable or because they do not propagate ($w_q=w_p=0$). 

For simplicity, we again restrict the discussion to $\gamma_1=\gamma_2=\gamma$.
For a fixed value of $\gamma$, the solutions of Eqs.(\ref{5}) are characterized
by all integers such that $|w_i|\leq[\gamma]$, where the square brackets mean
integer part \cite{lm}. For each possible set $(w_q,w_p)$ of winding numbers
there are four different solutions in $[0,1[^2$: $(q_0, p_0)$, $(q_0,
1/2-p_0)$, $(1/2-q_0, p_0)$ and $(1/2-q_0, 1/2-p_0)$, where
$q_0=\arcsin(-w_p/\gamma)/2\pi$ and $p_0= \arcsin(-w_q/\gamma)/2\pi$. For
$0<\gamma<1$ only $(w_q,w_p)=(0,0)$ is a solution and as a consequence there
are no period-$1$ accelerator modes in this parameter range since these orbits
do no propagate. As $\gamma$ increases, at each integer value there is a
first-order bifurcation (fold catastrophe) where several new orbits appear.
Taking into account all possible new couples of winding numbers and their signs
(where at least one is equal to $[\gamma]$ while the other one can be less than
$[\gamma]$), there are at each integer value of $\gamma$ a total number of
$32[\gamma]$ new orbits appearing. This makes, for a given value of $\gamma$, a
total number of $(4[\gamma]+2)^2$ period-$1$ fixed points. 

These orbits are stable if $|Tr {\bf M}|\leq 2$ and unstable otherwise, where
${\bf M}$ is the monodromy matrix. For a given couple $(w_q,w_p)$ the trace is 
given by
\begin{equation}\label{6}
Tr {\bf M}=2-4 \pi^2 \epsilon_{qp}\sqrt{(\gamma^2-w_q^2)(\gamma^2-w_p^2)}
\end{equation}
with $\epsilon_{qp}={\rm sgn}(\cos(2\pi q_0)\cos(2\pi p_0))$. At any integer
value of $\gamma$, $Tr {\bf M}=2$ for all the periodic orbits emerging at that
value of the parameter. $\epsilon_{qp}$ is positive for half of them, and
negative for the other half. For the latter half, it follows from (\ref{6})
that $Tr {\bf M}\geq 2 $ as $\gamma$ increases from the bifurcation point. On
the contrary, those having $\epsilon_{qp}=1$ are stable as $\gamma$ increases
from $[\gamma]$ but as soon as $Tr {\bf M}\leq -2$ they become unstable too.
This range of stability defines the domain of existence in parameter space of
period-$1$ accelerator modes. Its width is in fact relatively small (compare to
one), and its size decreases with $[\gamma]$. For example in the particular
case $w_q=w_p=w=[\gamma]$, $Tr {\bf M}= -2$ at $\gamma_{\rm c} =
([\gamma]^2+1/\pi^2)^{1/2} \approx [\gamma]+1/(2 \pi^2 [\gamma])$. 

At the value $\gamma_{\rm c}$ where $Tr {\bf M}= -2$ there is generically a
period doubling bifurcation, where a period-$2$ stable fixed point of the map
appears. If $\gamma$ further increases, this orbit will soon become unstable
through another bifurcation giving rise to the usual cascade of bifurcations.
We are not going to establish here a complete classification of the period-$2$
fixed points, but briefly discuss a special family who plays an important role
in the variations of the diffusion coefficient. These are the fixed points that
emerge at half integer values, $\gamma=m/2, \; m=1,3,\cdots$. Their point of
emergence is independent of $\gamma$ and given by $(q_0,p_0)=(1/4,0), (3/4,0),
(0,1/4)$ and $(0,3/4)$ with winding numbers $(w_q,w_p)=(0,-m), (0,m), (m,0)$
and $(-m,0)$, respectively. As for the period-$1$ orbits, their position and
stability varies with $\gamma$ and there is a small range in parameter space
close to $\gamma=m/2$ in which some of them are stable. Accelerator modes of
period $2$ therefore exist around all half integer values of $\gamma$. The
analysis of the accelerator modes can be pursued for arbitrary high periods,
with increasing complexity. 

\section{Deterministic diffusion}

From Eqs.(\ref{2}) it follows that
$$
p_n=p_0-\gamma_1 \sum_{i=0}^{n-1} \sin(2\pi q_i) 
$$
and then
$$
(\Delta p_n)^2\equiv (p_n-p_0)^2=\gamma_1^2\left[\sum_{i=0}^{n-1} \sin^2 (2\pi 
q_i) +2 \sum_{i=0}^{n-2} \sum_{j=i+1}^{n-1}  \sin (2\pi q_i)\sin (2\pi q_j)
\right] \ . 
$$
Our purpose is to compute the diffusion coefficient (\ref{3}) averaged over 
all possible initial conditions $(q_0,p_0)$ in the fundamental domain. By 
definition
$$
\langle f\rangle_{R_f} \equiv \int_0^1\int_0^1 dq_0 dp_0 f(q_0,p_0) \ .
$$
It is therefore implicit when computing any average value that 
the coordinates and momenta $(q_i,p_i)$ are expressed, using the 
equations of motion (\ref{2}), as a function of $(q_0,p_0)$. Because $\langle 
\sin^2 (2\pi q_i)\rangle=1/2$ and introducing 
the force-force correlation function
$$
C(\tau)=\langle\sin(2 \pi q_i) \sin(2 \pi q_{i+\tau}) \rangle_{R_f}
$$
it follows that
$$
\langle (\Delta p_n)^2 \rangle_{R_f} = \frac{\gamma_1^2}{2} n \left[ 1 + 4\sum_{\tau=1}
^{n-1} C(\tau)- \frac{4}{n}\sum_{\tau=1}^{n-1} \tau C(\tau) \right] \ .
$$
where we have used the invariance of $C(\tau)$ under time translations. If the 
$C(\tau)$ fall off rapidly enough with $\tau$, then \cite{cm}
\begin{equation}\label{7}
D =  \lim_{n\rightarrow\infty} \frac{\langle (\Delta p_n)^2 \rangle_{R_f}}{n} = 
\frac{\gamma_1^2}{2} \left[ 1 + 4\sum_{\tau=1}^\infty C(\tau) \right] \ .
\end{equation}
If correlations are neglected, i.e. the $q_i$ are considered as random 
uncorrelated variables, then it follows from Eq.(\ref{2}) that the motion in 
the $p$-direction reduces to a random walk with length-step $\gamma/\sqrt{2}$.
Accordingly, we obtain from Eq.(\ref{7}) $D=\gamma_1^2/2$. This is expected to 
be a good approximation for large values of the $\gamma$'s where the 
"randomization" is more effective. From this point of view, the corrections 
for $C(\tau), \, \tau=1,2,\cdots$ in Eq.(\ref{7}) can be considered as a 
large-$\gamma$ expansion.

The following relations are needed in order to compute the correlation 
functions
\begin{equation} \label{8}
\begin{array}{ccc}
\cos(z\sin\theta)&=&J_0(z)+2 \displaystyle \sum_{k=1}^\infty J_{2k}(z)\cos(2
k\theta)\\ 
\sin(z\sin\theta)&=&2 \displaystyle \sum_{k=0}^\infty J_{2k+1}(z)\sin((2
k+1)\theta) \ . 
\end{array}
\end{equation}
The one-step correlator is
$$ \begin{array}{ccl}
C(1)&=&\langle\sin(2\pi q_0) \sin(2\pi q_1)\rangle_{R_f} \\
    &=&\langle\sin(2\pi q_0) \sin\left[2\pi q_0+2\pi\gamma_2\sin(2\pi 
        p_0-2\pi\gamma_1\sin(2\pi q_0))\right]\rangle_{R_f}
\end{array}
$$
From Eqs.(\ref{8}) and simple trigonometric relations the averages can be 
computed explicitly. One finds
\begin{equation}\label{9}
C(1)=\frac{1}{2} J_0(2\pi\gamma_2) \ .
\end{equation}
The computation of the correlator for points differing by two time steps
$$ \begin{array}{ccl}
C(2)&=&\langle\sin(2\pi q_{-1}) \sin(2\pi q_1)\rangle_{R_f} \\
    &=&\langle\sin\left[2\pi q_0-2\pi\gamma_2\sin(2\pi p_0)\right]\sin\left[ 2 
\pi q_0+2\pi\gamma_2\sin(2\pi p_0-2\pi\gamma_1\sin(2\pi 
q_0))\right]\rangle_{R_f}
\end{array}
$$
follows the same lines, with increasing algebraic complexity. We get
\begin{equation}\label{10}
C(2)=\frac{1}{2} J_0^2 (2\pi\gamma_2) - \sum_{n=1}^\infty\left[ 
(-1)^{n+1} J_0 (2 n\pi\gamma_1) + J_2 (2 n\pi\gamma_1)\right]
J_n^2 (2 \pi\gamma_2) \ .
\end{equation}
From the computation of $C(3)$ we only keep the term $J_0^3(2 \pi\gamma_2)/2$,
the other being of higher order in powers of Bessel functions. Then, from
Eqs.(\ref{9}) and (\ref{10}) we finally have 
\begin{equation}\label{11}
\begin{array}{ccl}
D/(\gamma_1^2/2)&\approx& 1+2 \left[J_0 (2 \pi\gamma_2)+ J_0^2 
(2 \pi\gamma_2)+J_0^3 (2 \pi\gamma_2)\right] \\ & &-4 \displaystyle
\sum_{n=1}^\infty\left[ (-1)^{n+1}  J_0 (2 n\pi\gamma_1) +  J_2 (2 \pi n
\gamma_1)\right]  J_n^2 (2 \pi\gamma_2) \ . 
\end{array}
\end{equation}
In order to compare this result with the actual properties of the transport, we
have numerically computed the quantity $D/(\gamma^2/2)$ as a function of
$\gamma=\gamma_1=\gamma_2$. For each value of $\gamma$, the diffusion
coefficient was calculated from the slope of $\frac{\langle (\Delta p_n)^2
\rangle}{(n\gamma^2/2)}$ for $n=0,\cdots,3000$ averaged over $9000$ different
initial conditions taken at random on the elementary cell. The values obtained
are plotted in Fig.3, together with the analytical result (\ref{11}). The
approximation (\ref{11}) clearly accounts for the main features of the
diffusion process. Indeed, for most values of $\gamma$, the numerical points
are in very good agreement with the finite value obtained in (\ref{11}). 
The oscillating behaviour of period $\Delta\gamma=1$ can
be physically related to the strong modifications induced in the transport
properties by the appearance of period-$1$ accelerator modes at each integer
value of $\gamma$. For large values of $\gamma$ the oscillations are more
regular with decreasing amplitude $\propto 4/\pi \sqrt{\gamma_2}$. For small
values of $\gamma$ the phase space structure is more rich and accordingly $D$
is less regular. In this regime the agreement with Eq.(\ref{11}) is less good,
as shown in the inset of Fig.3, and below $\gamma\approx 0.35$ the theoretical
value is definitely wrong (indeed, from Eq.(\ref{11}) we get
$D/(\gamma/2)|_{\gamma=0} =7$, while the numerical values tend to $0$). To
compute more accurately the diffusion coefficient in this limit, we make use of
the results obtained in \cite{lw}. In the limit $\gamma \ll 1$ the diffusion
coefficient of the stochastic web takes the form 
\begin{equation} \label{12}
D_{\rm stoch} \approx \frac{2 \gamma^3}{\pi+2\gamma\left[3-\ln (c\pi)+ 3
\ln (\pi\gamma)\right]} \ , \;\;\;\;\;\;\; \gamma \ll 1
\end{equation}
where $c\approx 20$. Because our computations are averaged over all possible 
initial conditions, we shall multiply Eq.(\ref{12}) by the relative area ${\cal 
A}$ of the stochastic web \cite{lw}
$$
{\cal A} \approx \frac{c}{\pi^2 \gamma^3} {\rm e}^{-\pi/2\gamma} 
\;\;\;\;\;\;\; \gamma \ll 1
$$
with the result
\begin{equation} \label{13}
\frac{D}{(\gamma^2/2)} \approx \frac{4 c}{\pi^2}\frac
{{\rm e}^{-\pi/2\gamma}}{\gamma^2\left\{\pi+2\gamma\left[3-\ln (c\pi)+ 3
\ln (\pi\gamma)\right]\right\}} \ , \;\;\;\;\;\;\; \gamma \ll 1 \ .
\end{equation}
This function is plotted at the origin in the inset of Fig.3, and is in very 
good agreement with the numerical values.

On the other hand, the numerical results do not agree with the theory close to
low integer and half integer values of $\gamma$, where some of the numerical
points lie systematically higher. If the number of iterations employed in order
to compute the slope is increased, these points are not stable but rather move
up, a fact clearly pointing out towards a divergence of the diffusion
coefficient. Analytically, the divergence should come out from further
correlation terms in the expansion (\ref{7}) \cite{cm}. As mentioned in the
introduction, this divergent behaviour is closely related to the appearance of
accelerator modes. Because of their stability, these ballistic modes occupy a
(small but) finite phase-space volume. It then follows that $\langle
(p_n-p_0)^2\rangle_{\cal S} \sim n^2$ and $D\rightarrow\infty$ as $n$ when
$n\rightarrow\infty$ since the average was computed over the whole fundamental
domain (cf Eq.(\ref{3p})). For large integer or half integer values of $\gamma$
the divergence is not visible numerically because the area and range of
existence of the accelerator modes is very small (cf Section II). 

One could think that the divergence of the transport coefficient is limited to
stable regions of phase space where accelerator modes exist. This in fact is
not true. When this modes are present, two rather different dynamics coexist.
On the one hand, there is the ballistic motion associated with the small stable
regions. On the other hand there is the chaotic sea (occupying most of
phase-space as soon as $\gamma > 0.5$), where incoherent motion takes place and
where we may expect (as in the absence of accelerator modes) a regular
Brownian-type diffusion. But the frontier between both types of motion is not
sharp. Stable islands are surrounded by broken tori (called cantori). These
broken tori play the role of partial barriers for the motion in the chaotic
sea. When the particle is trapped by a cantorus in the neighborhood of the
accelerator mode, it is dragged across the phase plane ballistically. These
rare events can however dominate the dynamics and produce an anomalous
diffusion \cite{szk,zk}. If the trapping time probability distribution has the
asymptotic behaviour 
$$
\psi(n) \approx 1/n^{4-\beta}
$$
then we can classify the dynamics into three distinct regimes \cite{zk}
\begin{equation} \label{14}
\langle (\Delta p_n)^2 \rangle \sim \left\{\begin{array}{ccl}
n^2 & 2<\beta<3 & {\rm ballistic} \\ n^{\beta} & 1<\beta<2 & {\rm L\acute{e}vy} 
\\ n & \beta<1 & {\rm Brownian} \end{array} \right.
\end{equation}
The presence of accelerator modes may then induce strong modifications into the
transport properties of the chaotic component. This happens when the exponent 
$\beta$ lies between $1$ and $2$. The anomalous trajectories consists of a 
regular diffusion process punctuated by intermittent ballistic episodes. 
We find that this type of transport occurs in the kicked Harper map in certain 
parameter ranges. As an example, in Fig.4 we have plotted $5000$ iterations of 
a trajectory lying in the chaotic component obtained for $\gamma=1.011$. 
The orbit is shown in the entire phase plane as well as in the fundamental
domain. The former picture stresses the ballistic L\'evy-type intermittent 
behaviour of the orbit, while the latter makes clear the trapping of the orbit
around the accelerator modes. In fact, the long step propagating coherently
over more than $150$ fundamental domains is clearly visible in part (b) of that
figure as a concentration of points around the stable region associated with
the period-$1$ orbit emerging at $\gamma=1$ at $(q,p)=(1/4,3/4)$. 

In order to make more quantitative these considerations, we have 
numerically computed the exponent $\beta$ as a function of $\gamma$ for the 
chaotic component of phase space. This was done propagating for each value of 
$\gamma$ $9000$ trajectories iterated $5000$ times. The average over the
trajectories was taken in order to compute from Eq.(\ref{4}) the exponent 
$\beta$. The initial conditions of these trajectories were taken at random in a 
small region of size $0.05$ around the phase-space point $(0,1/2)$. This choice 
ensures that the orbits lie in the chaotic component of phase space. The 
results obtained are plotted in Fig.5 for $0.2<\gamma<2$, where it is clearly 
visible that values of $\beta$ greater than one are obtained for values of 
$\gamma$ slightly higher than $0.5$ and $1$. This confirms the existence of 
anomalous diffusion. However, it is remarkable that in spite of the fully 
developed fractal structure of phase space existing for $0<\gamma<1$, we 
observe anomalous diffusion only in a very localized region around $\gamma
\approx 0.5$. Could be that looking with a smaller step in $\gamma$ other peaks
will emerge. In Fig.6 the two peaks of Fig.5 are amplified. The function
$\beta(\gamma)$ in the second peak is a strongly oscillating function, and for
$\gamma \approx 1.05$ the anomalous diffusion disappears. This value is in good
agreement with the value $\gamma_{\rm c}\approx 1.051$ obtained in Section II
for the disappearance of the (last) period-$1$ accelerator mode emerging at
$\gamma=1$. 

\section{Conclusion}

The normal diffusion process of a deterministic two-dimensional area-preserving
map can thus be very well described in terms of a truncated expansion of the
diffusion coefficient. The theoretical prediction works well even in parameter
ranges where the fractal phase-space structure of the classical dynamics is
strong (mixed regime). However, when stable periodic orbits propagating
coherently through the chaotic sea are present, the diffusion coefficient
diverges. This divergence is trivial in the sense that it is directly related
to the existence of a finite and stable ballistic region of phase space. It is
not accounted for by the truncated expansion but should be recovered if
additional terms are added \cite{cm}. 

The presence of stable coherent orbits may also modify the transport properties
of the chaotic sea itself. We find that in certain parameter regions
surrounding the emergence of accelerator modes the diffusion is anomalous, with
$1<\beta<2$ in Eq.(\ref{4}) (and (\ref{14})). The puzzling fact is that this is
not always true. For example, we were not able to detect any anomalous
diffusion for parameter values close to (and slightly higher than) $\gamma =
1.5,\, 2.5, \cdots$, where we know that period-$2$ accelerator modes emerge,
but we do observed anomalous diffusion at $\gamma =0.5,\, 1,\, 2,\, 3, \cdots$.
This strengthens the need to develop a quantitative and predictive theory for
the anomalous diffusion in Hamiltonian systems, which does not exist today.

\pagebreak

\large 
\begin{center} 
FIGURES
\end{center} 
\normalsize 

\begin{description}

\item{FIG. 1:} The kicked Harper map plotted in the fundamental domain. (a) 
$\gamma=0.00063$, (b) $\gamma=0.31$, (c) $\gamma=0.565$ and (d) $\gamma=0.63$.

\item{FIG. 2:} Same as in Fig.1(b) but plotted in the full phase plane.

\item{FIG. 3:} Normalized diffusion coefficient as a function of 
$\gamma$. Analytical curve (solid line); numerical results (dots). Inset: 
amplification in the range $0<\gamma<2$; the full curve at the origin 
represents Eq.(\ref{13}).

\item{FIG. 4:} Chaotic trajectory obtained for $\gamma=1.011$ where anomalous 
diffusion occurs. (a) phase plane representation; (b) same trajectory as in 
part (a) but plotted in the fundamental domain.

\item{FIG. 5:} The index $\beta$ of Eq.(\ref{4}) as a function of $\gamma$ 
(dots). The interpolating full line is to guide the eye.

\item{FIG. 6:} Amplification of the two peaks of Fig.5. (a) Peak around 
$\gamma=0.5$. (b) Peak around $\gamma=1$.

\end{description}

\begin{figure}[p]
\epsfig{file=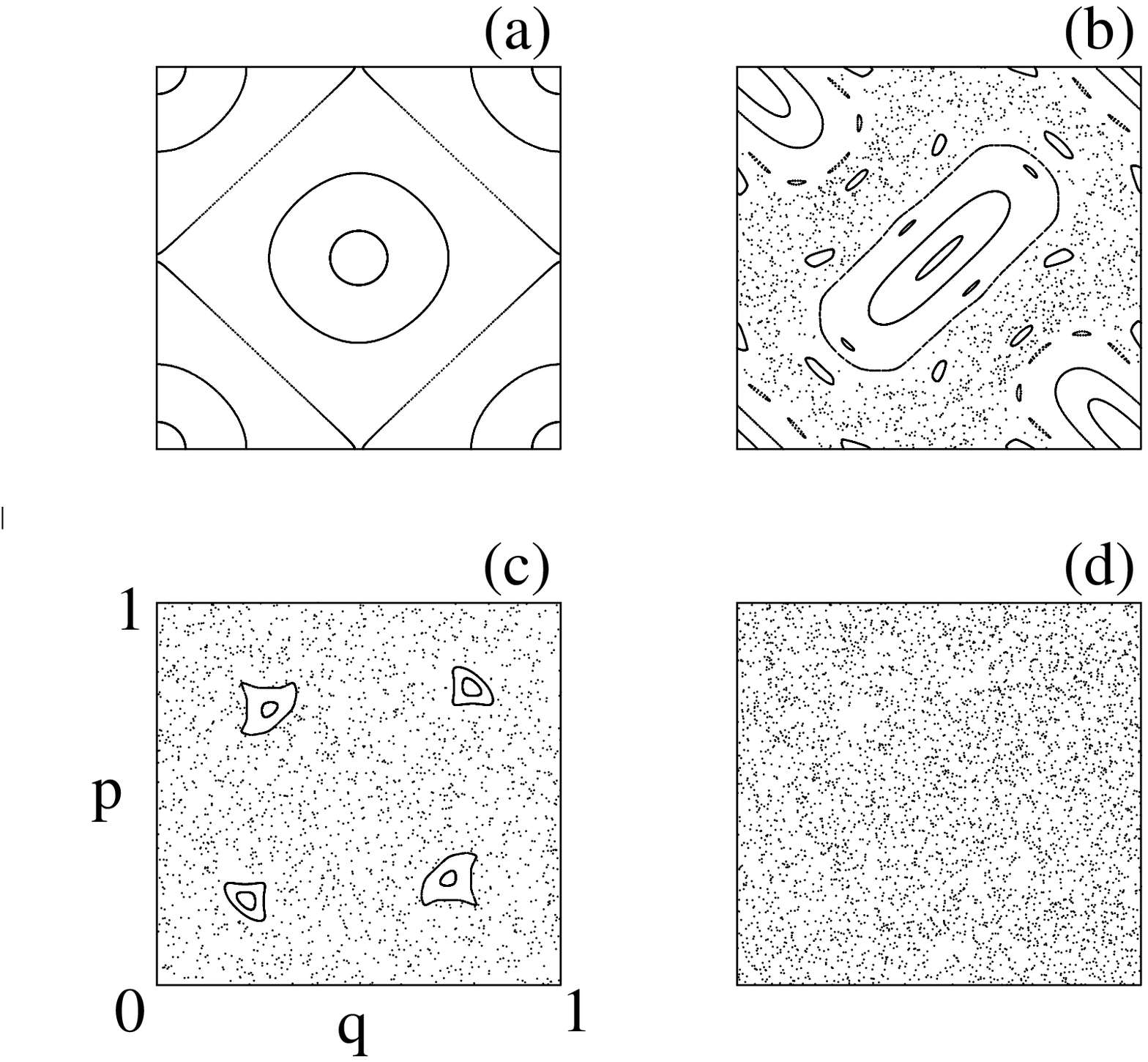}%
\end{figure}

\begin{figure}[p]
\epsfig{file=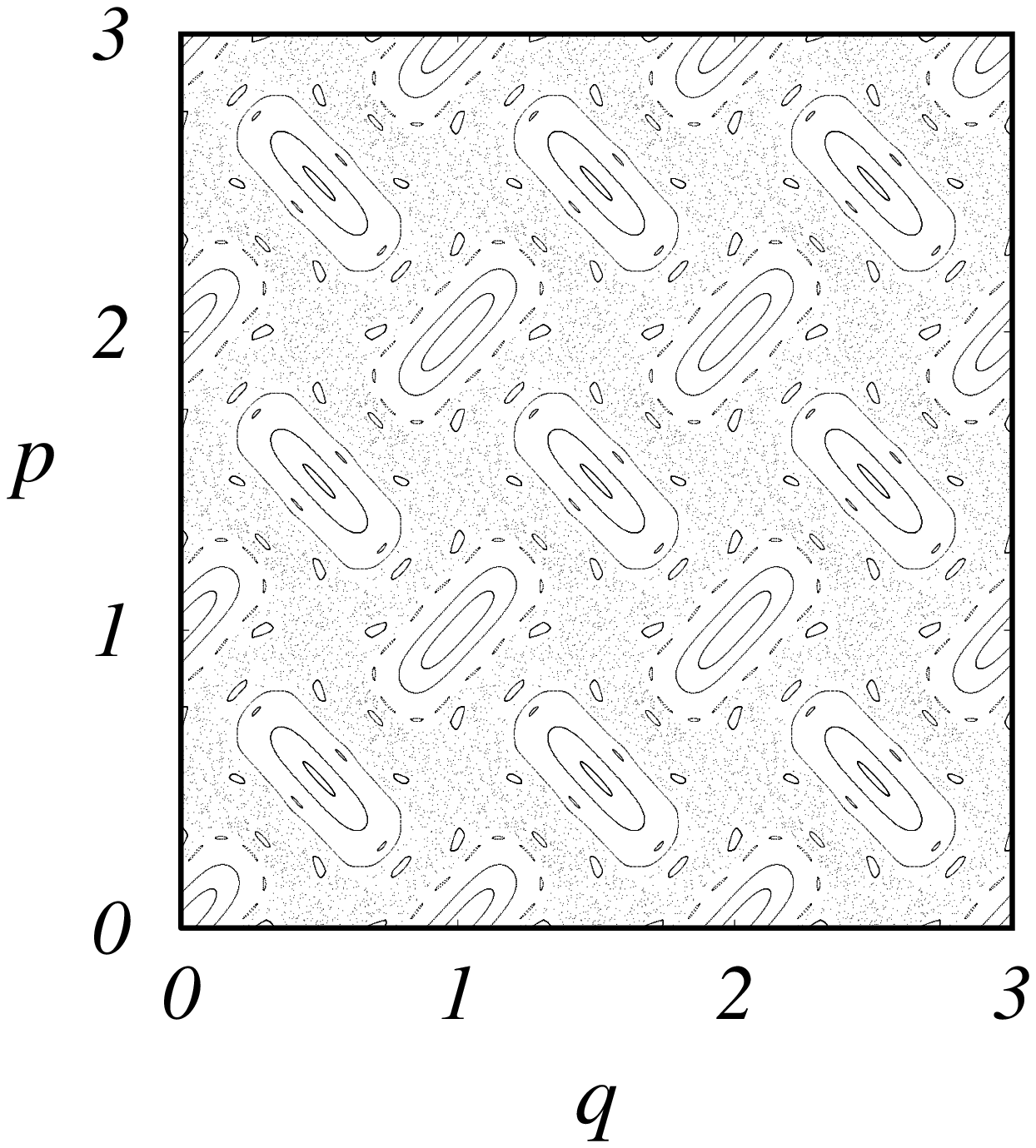}
\end{figure}

\begin{figure}[p]
\hspace{15cm}\makebox[1mm][r]{\epsfig{file=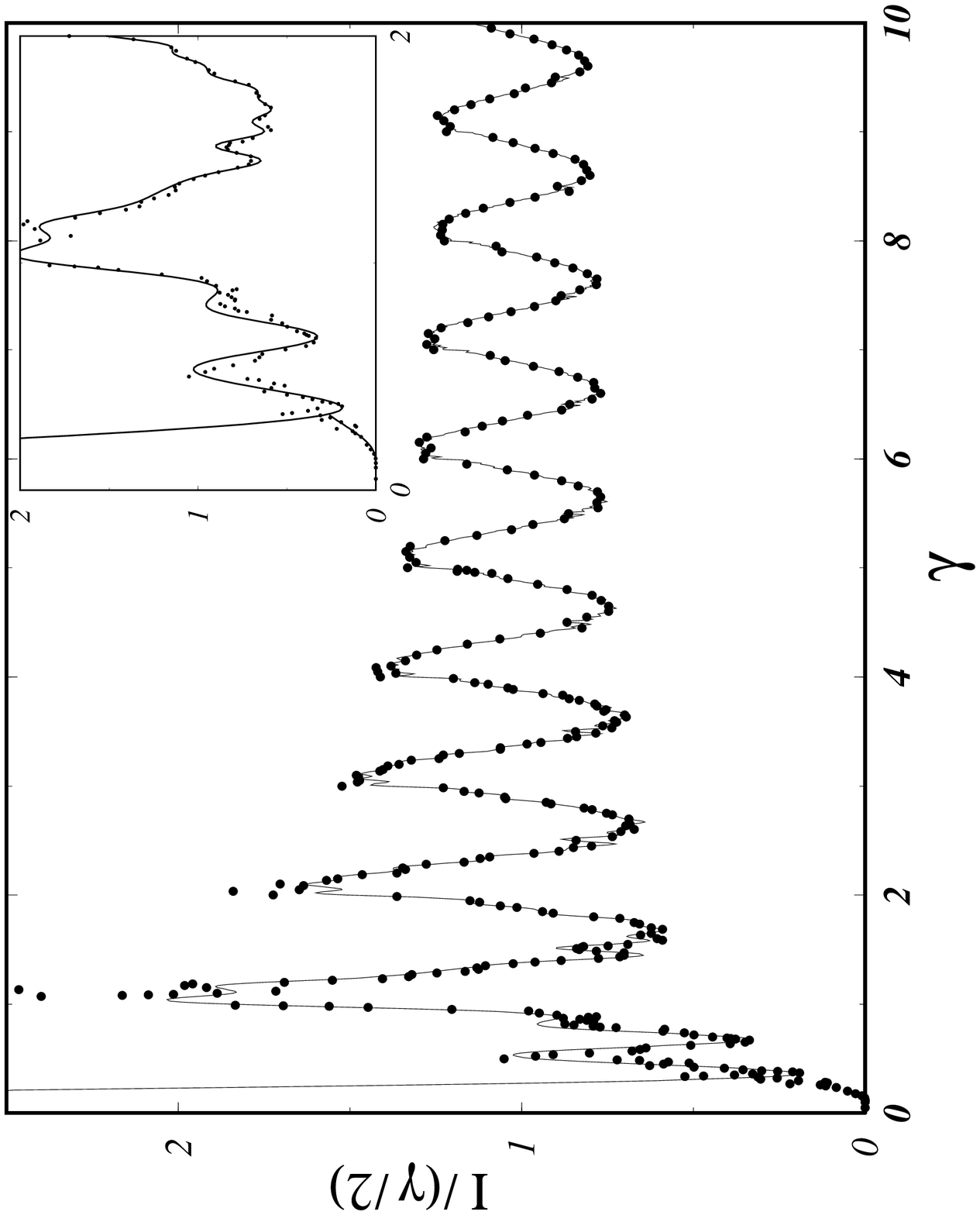}}
\end{figure}

\begin{figure}[p]
\epsfig{file=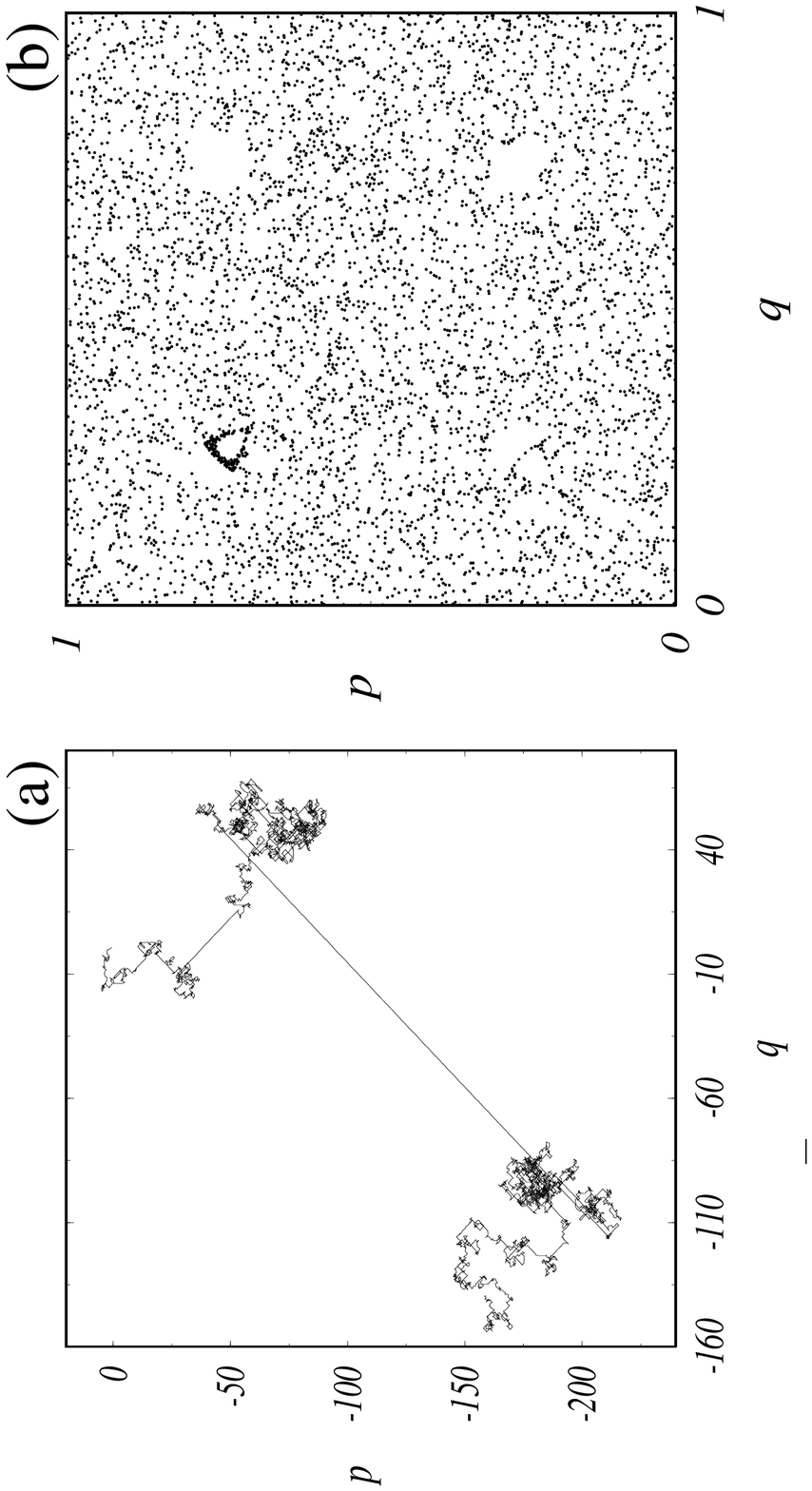}
\end{figure}

\begin{figure}[p]
\hspace{14.8cm}\makebox[1mm][r]{\epsfig{file=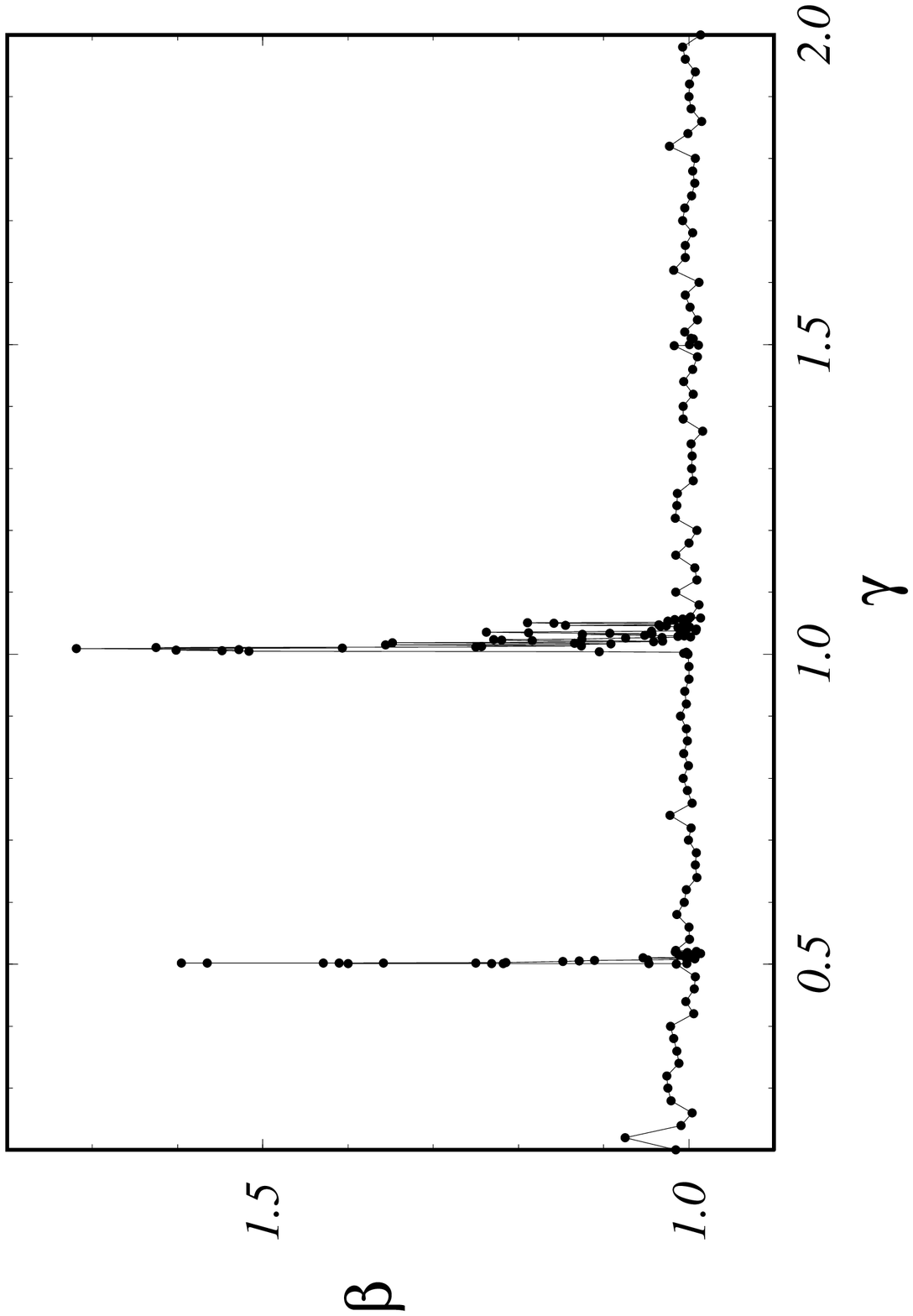}}
\end{figure}

\begin{figure}[p]
\epsfig{file=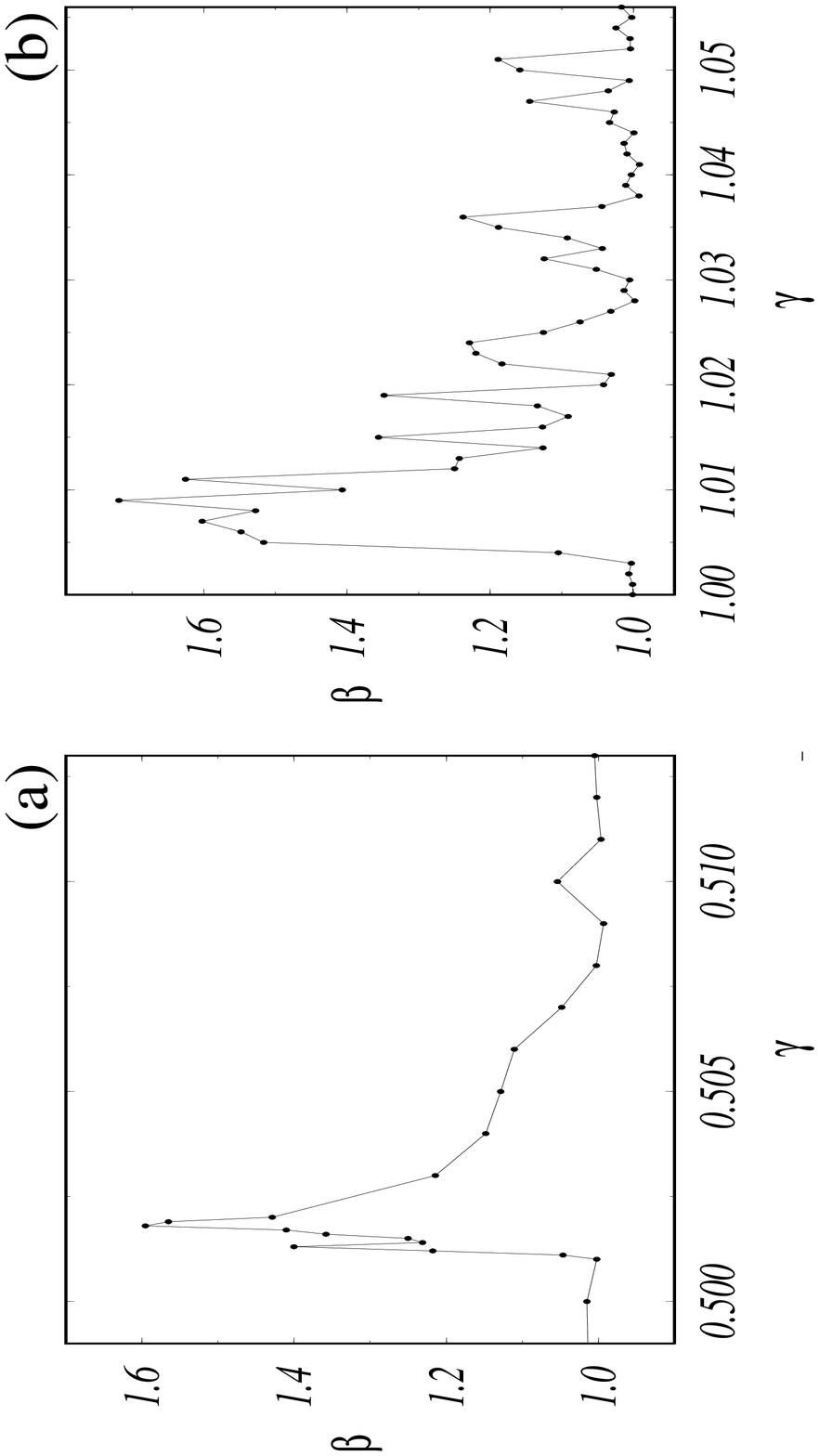}
\end{figure}

\end{document}